\documentclass[journal=jpccck,manuscript=article]{achemso}

\usepackage{chemformula} 
\usepackage[T1]{fontenc} 
\usepackage{color}
\usepackage{amssymb,amsmath}
\usepackage{dcolumn}
\usepackage{multirow}
\usepackage{hyperref}
\usepackage{siunitx}

\author{Quinn Campbell}
\affiliation[Sandia National Laboratories]{Center for Computing Research, Sandia National Laboratories, Albuquerque NM 87185}
\email{qcampbe@sandia.gov}
\author{Jeffrey A. Ivie}
\author{Ezra Bussmann}
\author{Scott W. Schmucker}
\affiliation[Sandia National Laboratories]
{Sandia National Laboratories, Albuquerque NM 87185}
\author{Andrew D. Baczewski}
\affiliation[Sandia National Laboratories]{Center for Computing Research, Sandia National Laboratories, Albuquerque NM 87185}
\author{Shashank Misra}
\affiliation[Sandia National Laboratories]
{Sandia National Laboratories, Albuquerque NM 87185}
\email{smisra@sandia.gov}

\title
  {A model for atomic precision p-type doping with diborane on Si(100)-2$\times$1}

\begin{document}

\begin{tocentry}
\includegraphics{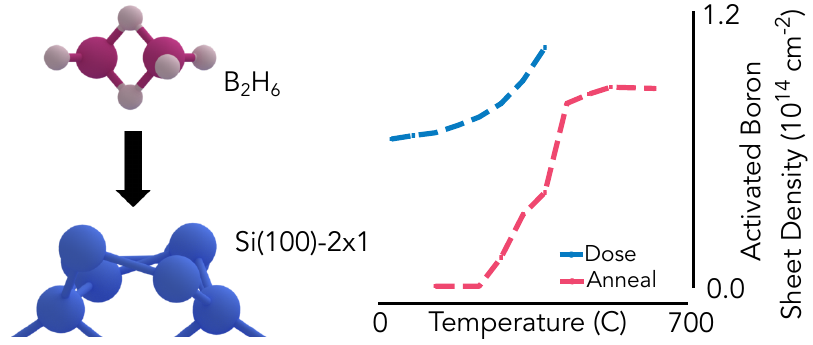}
\end{tocentry}

\begin{abstract}
Diborane (B$_2$H$_6$) is a promising molecular precursor for atomic precision p-type doping of silicon that has recently been experimentally demonstrated [T. {\v{S}}kere{\v{n}}, \textit{et al.,} Nature Electronics (2020)]. 
We use density functional theory (DFT) calculations to determine the reaction pathway for diborane dissociating into a species that will incorporate as electrically active substitutional boron after adsorbing onto the Si(100)-2$\times$1 surface.
Our calculations indicate that diborane must overcome an energy barrier to adsorb, explaining the experimentally observed low sticking coefficient ($<$ \SI{1e-4}{} at room temperature) and suggesting that heating can be used to increase the adsorption rate.
Upon sticking, diborane has an $\sim 50\%$ chance of splitting into two BH$_3$ fragments versus merely losing hydrogen to form a dimer such as B$_2$H$_4$.
As boron dimers are likely electrically inactive, whether this latter reaction occurs is shown to be predictive of the incorporation rate.
The dissociation process proceeds with significant energy barriers, necessitating the use of high temperatures for incorporation.
Using the barriers calculated from DFT, we parameterize a Kinetic Monte Carlo model that predicts the incorporation statistics of boron as a function of the initial depassivation geometry, dose, and anneal temperature.
Our results suggest that the dimer nature of diborane inherently limits its doping density as an acceptor precursor, and furthermore that heating the boron dimers to split before exposure to silicon can lead to poor selectivity on hydrogen and halogen resists.
This suggests that while diborane works as an atomic precision acceptor precursor, other non-dimerized acceptor precursors may lead to higher incorporation rates at lower temperatures.
\end{abstract}


\section{Introduction} 
\label{sec:introduction}
\begin{figure}
\includegraphics[width=0.6\columnwidth]{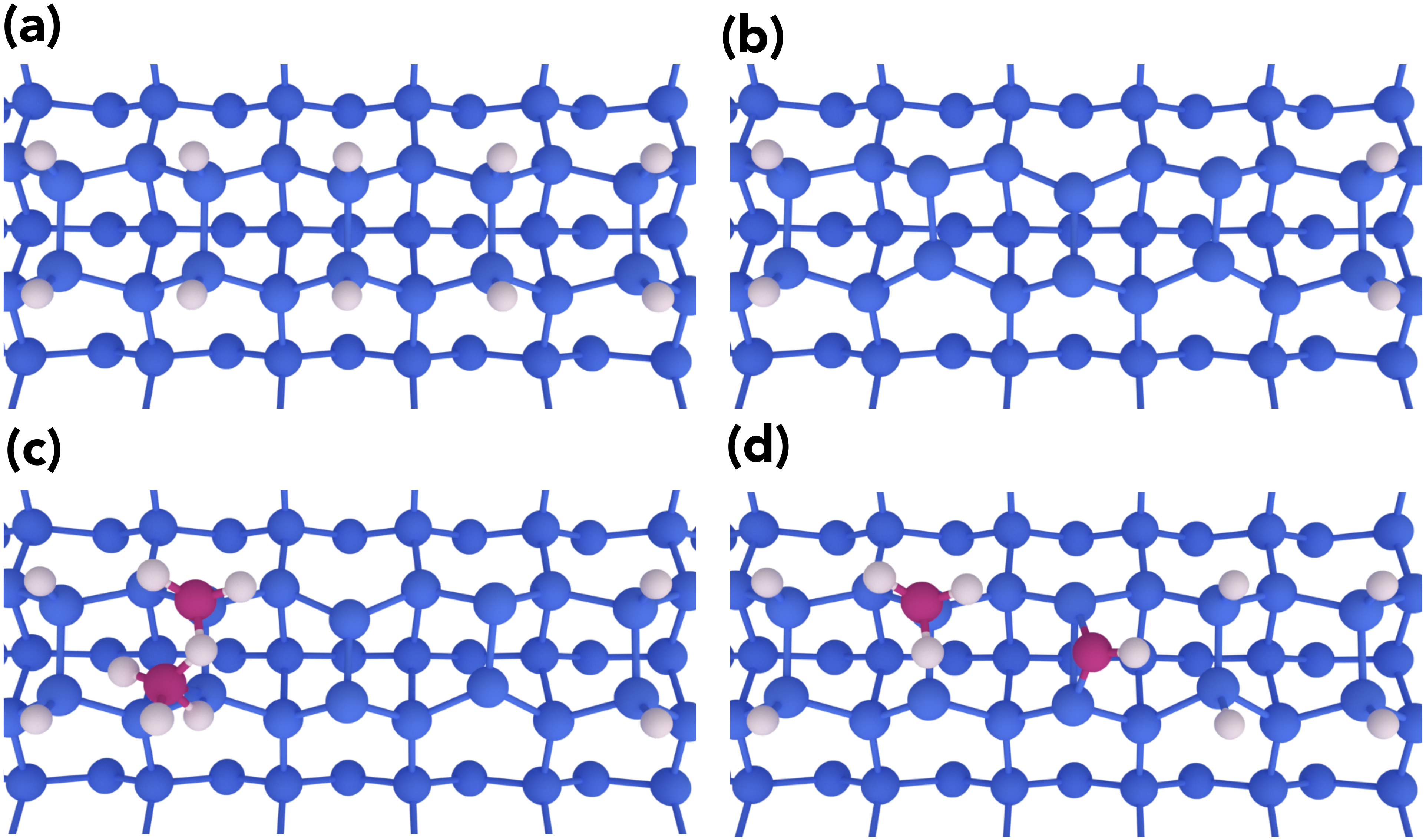}
\caption{
An outline of atomic-precision incorporation of electrically active boron from a diborane precursor.
(a) We start with a hydrogen-passivated Si(100)-2$\times$1 surface, where the hydrogen acts as a resist to the adsorption of a diborane molecule. 
(b) Using STM or photolithography, a ``window'' is opened in the silicon surface through hydrogen depassivation, exposing reactive dangling bonds.
(c) The silicon is exposed to diborane, shown here in its initial adsorption state, corresponding to configuration A1 in Fig.~\ref{fig:rxn_pathway_schematic}. 
(d) The diborane goes through a series of chemical reactions to arrive at a bridging configuration that will incorporate as a single electrically active substitutional boron atom in silicon. 
This bridging BH dimer corresponds to configuration C1 in Fig.~\ref{fig:rxn_pathway_schematic}. 
\label{fig:apam}}
\end{figure}

Atomic precision advanced manufacturing (APAM) promises to greatly improve capabilities for nanoelectronic device design, with the potential for realizing exotic quantum and classical devices that require a high-level of precision in the placement of dopants within silicon and/or dopant densities beyond the solid-solubility limit.~\cite{ward2020atomic} 
APAM begins with a Si(100)-2$\times$1 surface that is terminated with a single monolayer of resist atoms.
Most prominently, a monohydride termination (see Figs.~\ref{fig:apam}a) has been used, but halogen resists are an active area of research as well.~\cite{pavlova2018first,dwyer2019stm,silva2020reaction}
A tool such as a Scanning Tunneling Microscope (STM)\cite{lyding1994nanoscale} or a pulsed UV laser~\cite{katzenmeyer2020photothermal} is then used to selectively remove the resist atoms from a region of interest (see Figs.~\ref{fig:apam}b), after which the surface is exposed to a precursor molecule containing the desired dopant atom.
The precursor gas selectively adsorbs onto the bare silicon as opposed to the surrounding resist (see Figs.~\ref{fig:apam}c) and, through a series of chemical reactions, leads to a bridging configuration (see Figs.~\ref{fig:apam}d) that will eventually lead to an incorporated dopant atom.
For a phosphine (PH$_3$) precursor gas, a phosphorus donor atom can be placed onto a silicon surface to within one lattice site of a particular target.~\cite{Schofield2003}

While chemistries for atomic precision placement of donors have been well-developed both theoretically and experimentally,\cite{Schofield2003,Wilson2004,Warschkow2005,Wilson2006,Warschkow2016,pavlova2018first,wyrick2019atom,stock2020atomic} less work has been done in developing a similar chemistry for atomic precision acceptor placement.
However, such an advance is essential to developing processes that would enable the creation of acceptor-based quantum devices, p-n junctions, or more general integration with standard CMOS electronics.\cite{shim2014bottom,van2014probing,vskerevn2018cmos,vskerevn2020bipolar}
Among the key choices that need to be made are the chemical composition of the resist and the precursor molecule that reacts with the exposed surface to yield an atom that can incorporate in an electrically active configuration.
The precursor molecule for acceptor doping should selectively adsorb onto bare silicon relative to the resist and decompose into an incorporated acceptor atom with the addition of only a small amount of thermal energy. 
Diborane (B$_2$H$_6$) was recently demonstrated for p-type $\delta$-doping with a monohydride resist, serving as a relatively simple precursor for boron incorporation analogous to phosphine for phosphorus.~\cite{vskerevn2020bipolar}

In this paper, we theoretically study diborane as an APAM acceptor precursor on Si(100)-2$\times$1 with both a hydrogen resist, given its established selectivity for phosphine, and a chlorine resist, due to recent interest.\cite{dwyer2019stm,silva2020reaction}
We use Density Functional Theory (DFT) to calculate the reaction pathway for diborane decomposition on silicon surfaces, along with reaction barriers associated with moving between different configurations. 
We then utilize these reaction barriers and configurations in a Kinetic Monte Carlo (KMC) model that predicts the geometry and rate of incorporation.
We consider both small windows in the resist intended to incorporate one or a few boron atoms, as well as larger windows up to tens of nanometers wide.
By assuming that substitutional dimerized boron will be electrically inactive, we are able to reproduce incorporation rates consistent with experiment.
This leads us to conclude that the dimer nature of diborane limits its performance compared to the analogous, non-dimerized phosphine precursor.
 
\section{Methods}
\label{sec:methods}

\subsection{Electronic Structure Calculations}

We determine the thermodynamic adsorption energy of any particular configuration with the following equation:
\begin{equation}
	E_{\rm a} = E_{slab/adsorbate} - E_{slab} - E_{molecule},
\end{equation}
where $E_{\rm a}$ is the adsorption energy of the molecule on the silicon surface, $E_{slab/adsorbate} $ is the total energy of the adsorbate on the slab, $E_{slab} $ is the total energy of the slab without any adsorbate, and $E_{molecule}$ is the total energy of the isolated molecule. 
Negative values of $E_{\rm a}$ therefore imply a thermodynamically favorable adsorption energy for that configuration.
All total energy calculations are performed using the plane wave {\sc quantum-espresso} software package.\cite{Giannozzi2009} 
To compute reaction barriers between configurations we use the Nudged Elastic Band (NEB) method, also implemented in {\sc quantum-espresso}. 
We use norm-conserving pseudopotentials from the PseudoDojo repository~\cite{VanSetten2018} and the Perdew-Burke-Ernzerhof exchange correlation functional.~\cite{Perdew1996}
We use kinetic energy cutoffs of 50 Ry and 200 Ry for the plane wave basis sets used to describe the Kohn-Sham orbitals and charge density, respectively.
We use a 2$\times$2$\times$1 Monkhorst-Pack grid to sample the Brillioun zone.\cite{monkhorst1976special}

We perform all adsorption energy calculations on the 4$\times$4 supercell of a seven-layer thick Si(100)-2x1 slab with a 20 \AA~vacuum region. 
We place a hydrogen resist on the surface with the exception of three dimer sites, allowing us to gauge the selectivity of the diborane molecules on a bare silicon surface versus a resist-terminated surface. 
On the other end of the slab, the dangling bonds of the silicon are passivated with selenium atoms to prevent spurious surface effects. 
Selenium was determined to be optimal for achieving this purpose with minimal strain.
The bottom four layers of the slab are frozen and the geometry of the surface layers along with the adsorbate are relaxed until the interatomic forces are lower than 50 meV/\AA.
We compute the reference molecular energy for a single diborane molecule in a 15 \AA$^3$ box.

\subsection{Kinetic Monte Carlo}

We then use a Kinetic Monte Carlo (KMC) model \cite{Bortz1975,Gillespie1976} implemented using the KMCLib software package \cite{Leetmaa2014} to predict the incorporation rate of boron atoms in patches of exposed silicon that can be several nanometers wide. 
Our KMC uses transition rates based on the Arrhenius equation $ \Gamma = A \exp{\Delta/k_{\rm B}T}$,~\cite{Arrhenius1889}  where $\Gamma$ is transition rate, $A$ is the attempt frequency, $\Delta$ is the reaction barrier found from our earlier DFT calculations, $k_{\rm B}$ is the Boltzmann constant, and $T$ is the temperature. 
We set all attempt frequencies $A$ to $10^{12}$ s$^{-1}$ as a reasonable order of magnitude estimate based on an analysis of attempt frequencies for the dissociation of phosphine on silicon.\cite{Warschkow2016} 
We could use the Vineyard equation to compute more precise attempt frequencies within DFT, but given the simplicity of the pathway that we propose and the relatively low level of accuracy of the barrier height calculations, a simple estimate suffices to study the qualitative aspects of the incorporation chemistry.
Unless otherwise noted, we follow the incorporation schedule layed out by \v{S}kere\v{n} and coworkers,\cite{vskerevn2020bipolar} utilizing a dosing pressure of 1.5 $\times$ 10$^{-7}$ Torr for 10 minutes at \SI{120}{\celsius}, followed by an anneal at \SI{410}{\celsius} for 1 minute. 

We repeat each KMC calculation 200 times to obtain a meaningful statistical sampling of likely outcomes and report the average outcomes, along with standard deviations as applicable. 
We calculate the standard error by assuming a Poisson distribution of measured counts and using the standard error based on sample size. 

\section{Results}
\label{sec:results}

\subsection{Reaction Pathway}
\label{ssec:rxn_pathway}
\begin{figure*}[ht]
\includegraphics[width=\textwidth]{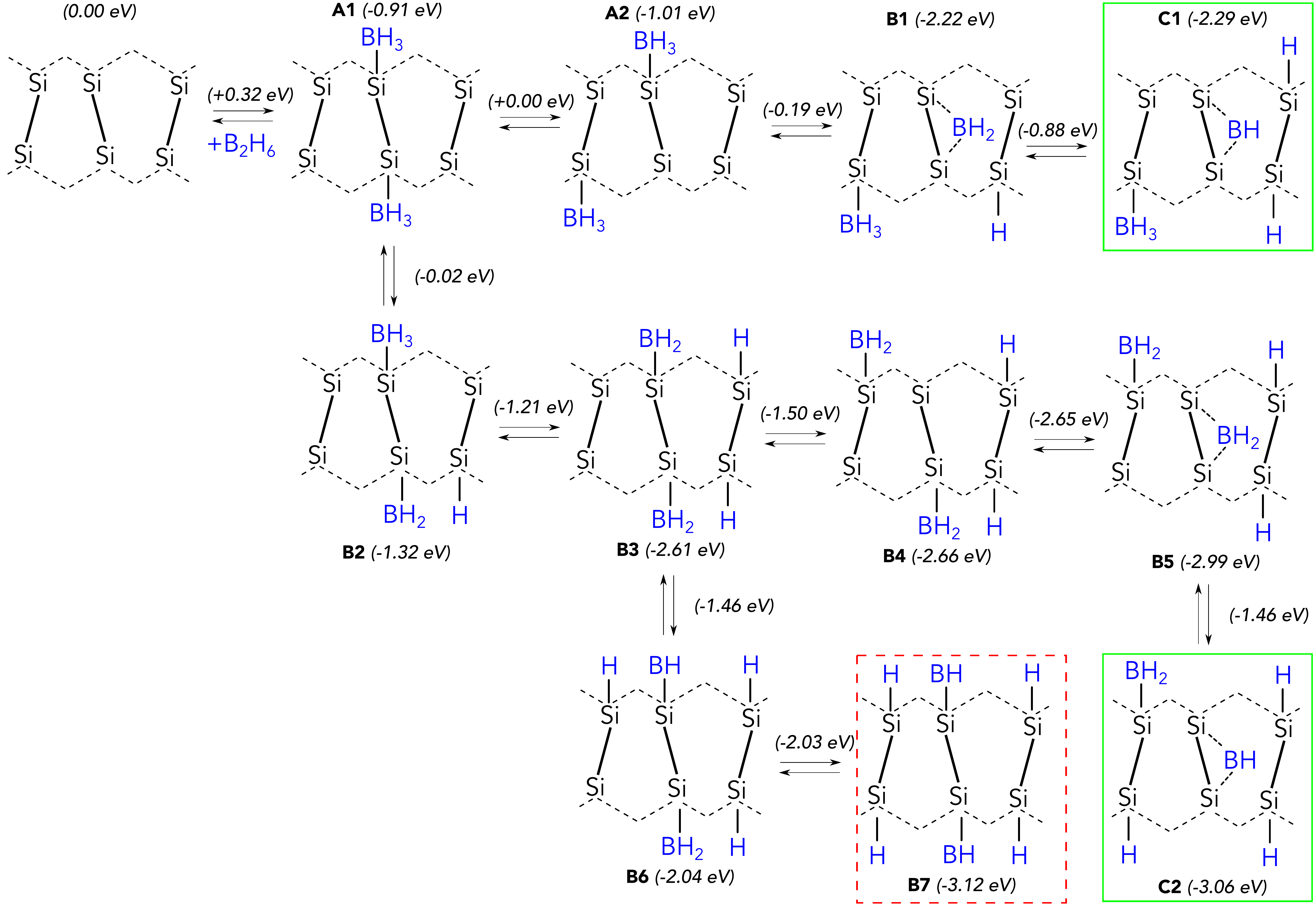}
\caption{The reaction pathway of diborane on a silicon surface. 
After surmounting an initial barrier for adsorption, the molecule has several potential dissociation pathways. 
We frame positive outcomes, where a bridging BH molecule is formed, with a green rectangle. 
We frame negative outcomes, where the two boron atoms remain on the same dimer leading to electrically inactive incorporation, with red dashed rectangles. 
The adsorption energy of each configuration is shown in parenthesis and by each reaction arrow is the adsorption energy of the corresponding transition state between the two configurations.  \label{fig:rxn_pathway_schematic}}
\end{figure*}

The reaction pathway of diborane on a silicon surface was determined using DFT, with the lowest energy configurations shown in Fig.~\ref{fig:rxn_pathway_schematic}. For ease of reference, we adopt a similar naming convention to the one developed by Wilson \textit{et al.} \cite{Wilson2004} Structures labeled with an A consist entirely of BH$_3$ fragments, structures labeled with a B consist of BH$_2$ fragments, and finally structures labeled with a C have bridging BH fragments. 

We predict diborane requires overcoming an energy barrier to adsorb into the lowest thermodynamic energy configuration, explaining the experimental observation that diborane has a low sticking coefficient ($<$\SI{1e-4}{}) at room temperature.\cite{yu1986doping,Wang1996stm} 
The lowest energy adsorption site for diborane on silicon requires a partial dissociation of the diborane molecule into two BH$_3$ fragments, each sitting on an opposite end of the dimer (configuration A1). 
This dissociation requires overcoming a reaction barrier of 0.32 eV to achieve an adsorption energy of --0.91 eV. While the ultimate thermodynamic stability of this adsorption is comparable to the reported --0.7 eV adsorption energy of phosphine,\cite{Warschkow2016} the addition of the reaction barrier implies that unless energy is provided to begin the dissociation process, diborane will not stick to the silicon surface. 
In contrast, on a hydrogen resist, diborane only weakly physisorbs with an adsorption energy of --0.02 eV and does not dissociate. We thus predict that heating the surface will lead to significantly better adsorption onto bare silicon, as demonstrated by \v{S}kere\v{n} \textit{et al.}\cite{vskerevn2020bipolar} 

The resulting pathway of the diborane fragments is then determined by whether the BH$_3$ fragments start shedding hydrogen to nearby dimers or whether the BH$_3$ fragments split to occupy different dimers. 
The pathway where diborane loses hydrogen to a nearby silicon dimer site (configuration B2) is more thermodynamically favorable with an adsorption energy of --1.32 eV as opposed to the configuration where the two BH$_3$ fragments split (A2), which has an adsorption energy of --1.01 eV. 
The reaction barriers for each path are essentially identical, however, with the A1 to B2 reaction having a barrier of 0.89 eV as opposed to the 0.91 eV barrier of the A1 to A2 reaction. 
Both of these barriers are relatively high, requiring higher temperatures than the initial adsorption reaction. 
A molecule in the A1 configuration will therefore have an approximately equal chance of moving toward A2 or B2. 
This reaction defines the entire resulting path of the diborane molecule, fully illustrated in Fig.~\ref{fig:rxn_pathway_schematic}. 
For the sake of limiting our computation to a feasible subset of possible configurations and in analogy to similar work on phosphine dissociation,\cite{Warschkow2016,schofield2006phosphine} we assume that once a diborane molecule reaches a bridging BH position, as in C1 and C2, it will eventually incorporate into the system. 

In contrast, if the two boron atoms remain on the same dimer, as in the B7 configuration, we assume that they will incorporate into the silicon lattice as a dimerized B$_2$ molecule which will likely be electrically inactive.
This is substantiated by prior work on boron clustering in silicon in which electrically inactive complexes may rationalize the presence of immobile boron in studies on ion implanted samples.~\cite{tarnow1992theory,stolk1995implantation,zhu1996ab}
Accordingly, the first step that limits the incorporation of electrically active boron is at A1, and whether the reaction proceeds to A2 or B2.
The next step that determines electrically active incorporation is at B3, where a dimer with a BH$_2$ molecule on both sides can either continue to lose hydrogen to a nearby dimer (B6), or split apart such that the BH$_2$ molecules move to separate dimers (B4).
This allows for non-dimerized bridging BH to form (C2). 

\begin{figure}
\includegraphics[width=0.6\columnwidth]{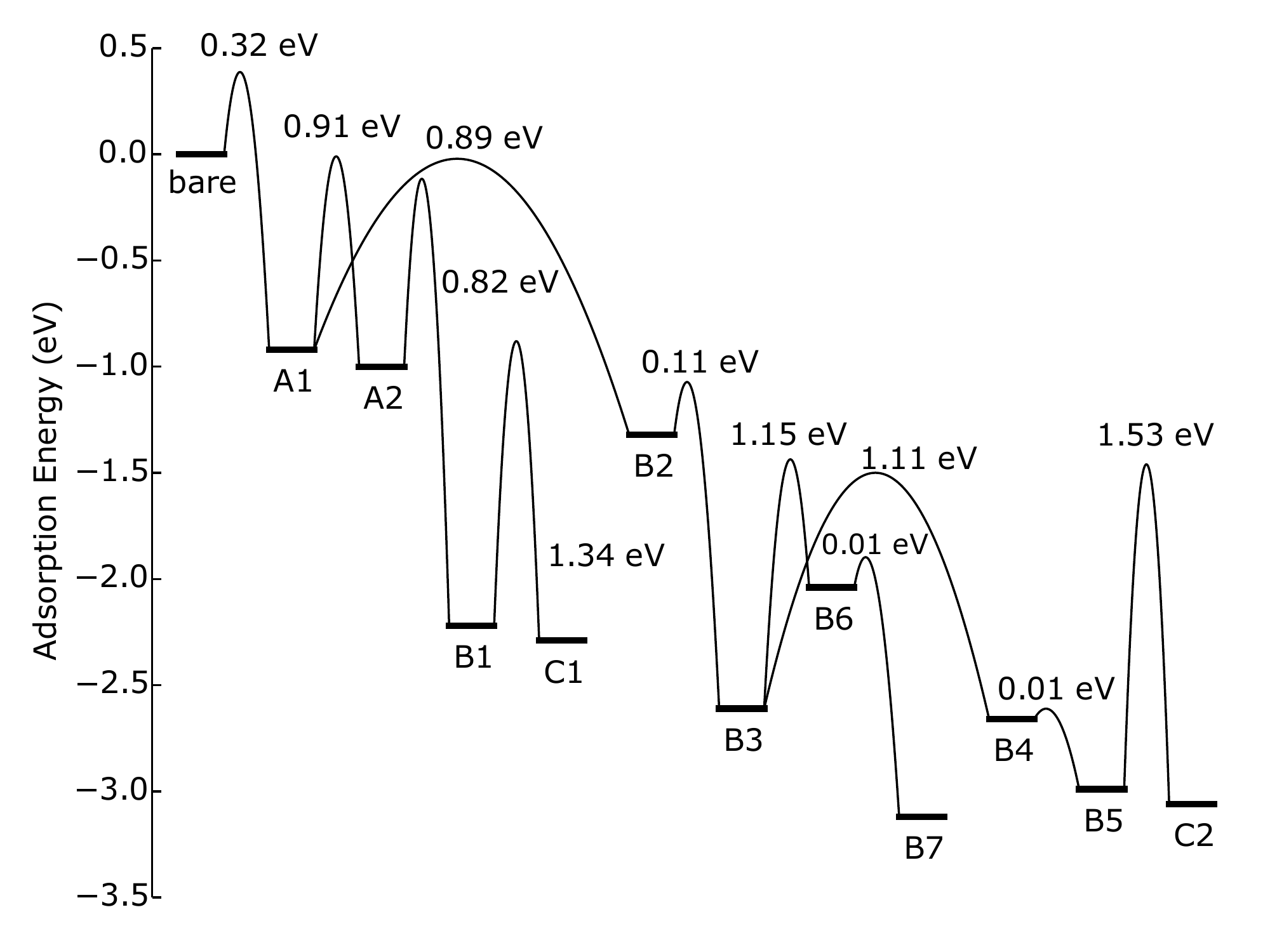}
\caption{Energetics pathway for diborane adsorbing onto a Silicon surface. Labels here refer to the configurations shown in Fig.~\ref{fig:rxn_pathway_schematic}. \label{fig:rxn_pathway_energetics}}
\end{figure}

\begin{figure*}
\includegraphics[width=\textwidth]{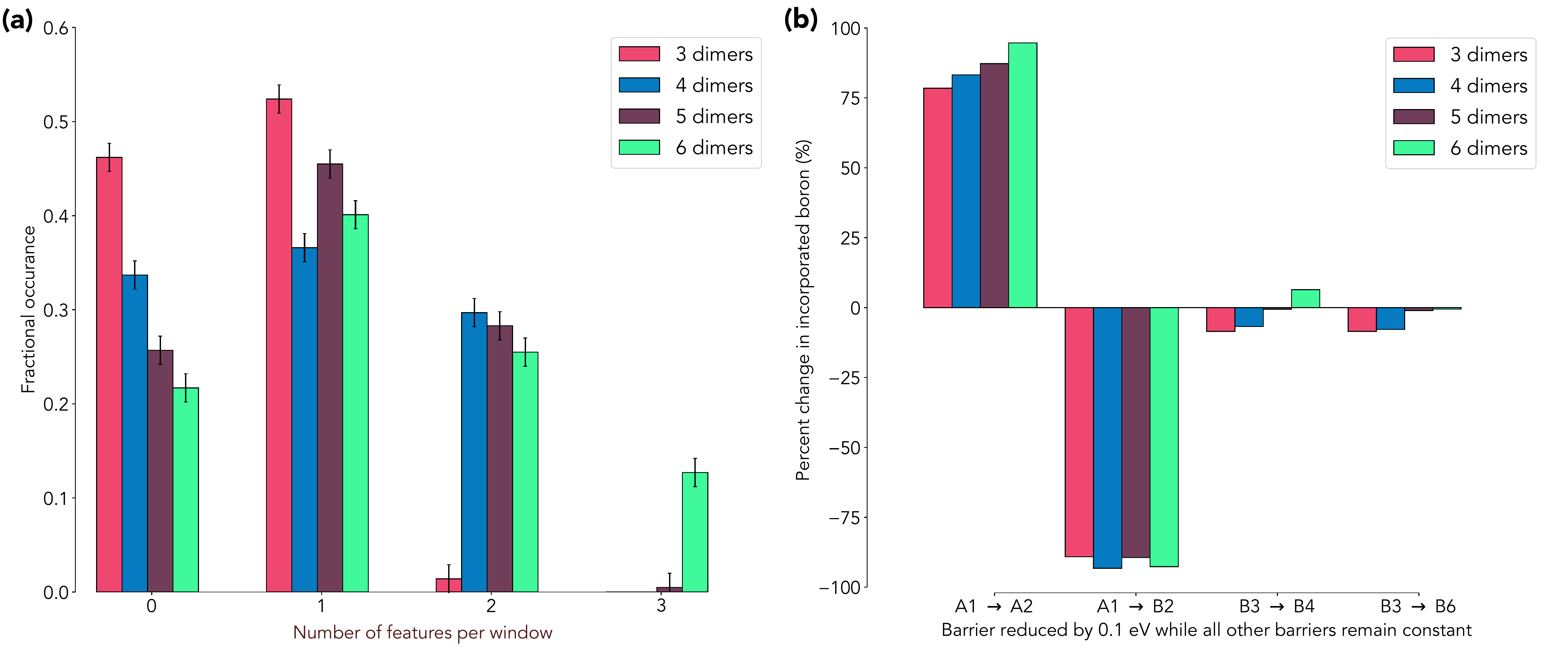}
\caption{ (a) Predicted rates of incorporation of a boron atom into a silicon window using Kinetic Monte Carlo simulations parameterized based on the reaction barrier shown in Fig.~\ref{fig:rxn_pathway_energetics}. (b) The percent change in the number of incorporated boron atoms while individual reaction barriers are lowered by 0.1 eV while holding all other barriers constant. The wide variation in expected incorporations based on changes to the A1 to A2 and A1 to B2 reaction barriers highlights the importance of the A1 configuration as a decision point for final incorporation.
\label{fig:dimer_window}}
\end{figure*}

The thermodynamic pathway for diborane dissociating and incorporating is entirely downhill, as shown in Fig.~\ref{fig:rxn_pathway_energetics}. Barriers remain high throughout the reaction process, however, with a dissociating molecule routinely having to overcome barriers on the order of $\sim$0.9 - 1.1 eV. The two incorporation reactions in particular, B1 to C1 and B5 to C2, have especially high barriers of $\sim$1.3 eV and $\sim$1.5 eV, respectively. These large reaction barriers necessitate high processing temperatures to result in significant numbers of incorporations. If we assume that an adsorbed diborane molecule will always take the lowest barrier option available to it, the molecule will eventually result in a bridging BH molecule in configuration C2. Two crucial configurations, however, A1 and B3, have two potential reactions with nearly identical reaction barriers, meaning molecules in these configurations will have nearly equal chances of which dissociation path they take, making the dissociation pathway of diborane highly stochastic.

\subsection{Incorporation Rate}
\label{ssec:incorporation}

To understand the impact of this dissociation pathway stochasticity on the final incorporation rate of diborane, we parameterize a Kinetic Monte Carlo (KMC) model with the reaction barriers calculated from DFT and displayed in Fig.~\ref{fig:rxn_pathway_energetics}. 
We initially measure the incorporation rate of diborane in small windows of silicon that have been depassivated with the hope of incorporating a single acceptor within the space of a few lattice sites. As shown schematically in Fig.~\ref{fig:rxn_pathway_schematic}, a diborane molecule needs a space at least three dimers wide to have sufficient room for the boron dimer to split and to lose hydrogen. We therefore measure the number of boron incorporation events in three, four, five, and six dimer wide windows in Fig.~\ref{fig:dimer_window}a. For ease of discussion in the following section, we introduce the notation $P_I(n|w)$  to describe the probability of $n$ acceptors incorporating into a $w$-dimer wide window. 

We find no dimer width that guarantees deterministic incorporation of a boron atom, although the probability of at least one incorporation event, $P_I(n >1 |w)$  does increase as a function of width $w$. A three dimer wide window has the highest probability of a single acceptor incorporation with $P_I(n=1|w=3) = 0.53$. This is lower than the equivalent probability of incorporation for a phosphine molecule, which has been experimentally measured at $\sim$ 0.7.\cite{Fuechsle2011} This lowered probability can largely be attributed to the larger size of the diborane molecule, requiring several more steps for dissociation and splitting of the boron dimer than a phosphine molecule. 

Because the rates in our KMC model are exponentially sensitive to barriers calculated using DFT, small changes in their values relative to the thermal energy can lead to changes in the final results of our analysis. 
For the temperatures relevant to our incorporation chemistry, the size of errors typical to DFT are sufficiently large that we should not take the quantitative predictions too seriously without considering the sensitivity of outcomes to errors $\mathcal{O}$(0.1 eV).
Accordingly we need to focus on particular steps in the incorporation for which the molecule has two nearly equal barriers that it must choose between (e.g. configurations A1 and B3). 
Slightly favoring the A1 to A2 reaction by reducing the A1 to A2 barrier by merely 0.1 eV can increase the incorporation rate dramatically, leading to $P_I(n=1|w=3) = 0.94$. In contrast, decreasing the A1 to B2 barrier by 0.1 eV, which favors the molecule losing hydrogen as opposed to splitting onto separate dimers, ensures virtually no incorporation with $P_I(n=1|w=3) = 0.05$. In contrast, the B3 configuration is revealed to be a less important decision point in the dissociation of diborane: reducing either the B3 to B4 reaction or the B3 to B6 reaction by 0.1 eV leads to insignificant changes in incorporation levels. 
This can be attributed to the fact that the reverse reactions of note, B4 to B3 and B7 to B6, have relatively low reaction barriers (1.16 eV and 1.15 eV, respectively) compared to the incorporation reaction of interest B5 to C2 with a barrier of 1.53 eV.
Thus a BH$_2$ cluster in the B4 configuration is, in the end, slightly more likely to return to the B3 configuration than to fully incorporate as an electrically active dopant, reducing the impact of whether the BH$_2$ molecules split to different dimer.
This emphasizes the importance of the A1 configuration and the resulting splitting (or lack thereof) of the boron dimer to determining the ultimate incorporation levels of diborane. 
This suggests future theoretical work to focus on increasing the accuracy of energies for this configuration and the associated reaction barriers. 

\begin{figure}
\includegraphics[width=0.6\columnwidth]{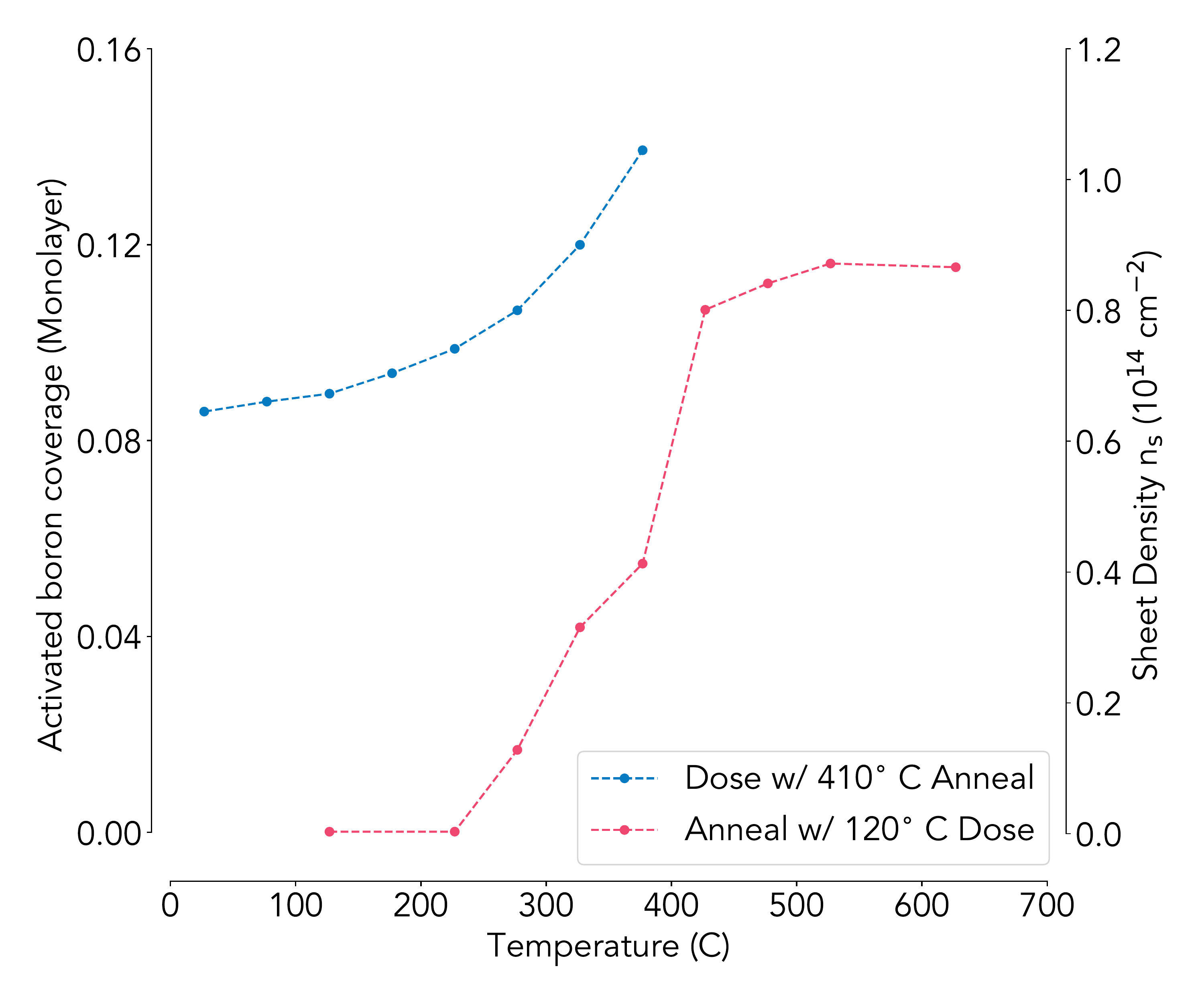}
\caption{ Predicted coverage and sheet density of activated boron dopants of a 10$\times$10 nm depassivated window where the dose temperature is varied while keeping the anneal temperature constant at 410 C or, alternately, varying the anneal temperature while keeping the dose temperature constant at 120 C.
\label{fig:temperature_dependence}}
\end{figure}

For $\delta$-doping, the final incorporation level of boron into silicon is highly dependent on both the dose and anneal temperature, as demonstrated by simulations of larger 10$\times$10 nm depassivated systems. 
In Fig.~\ref{fig:temperature_dependence}, we predict the monolayer coverage and sheet density of activated dopants in a system after being dosed with diborane while independently altering the dose and anneal temperature. 
The incorporation rate increases with dose temperature, although not significantly until reaching $\sim$\SI{250}{\celsius}. 
This can be attributed to the temperature becoming sufficiently high to activate the B1 to C1 reaction (with a barrier of 1.34 eV), which has a typical time scale (taken here as simply $1/\Gamma$ where $\Gamma$ is the Arrhenius rate of an equation) of $\sim$190 s at \SI{200}{\celsius}, but only $\sim$0.06 s at \SI{350}{\celsius}. 
Similarly, the B5 to C2 incorporation reaction becomes feasible only as  temperatures increase with a typical reaction time scale of $\sim$\SI{1.2e4}{s} at \SI{200}{\celsius} and $\sim$0.03 s at \SI{450}{\celsius}. 
It is likely that the ability for diborane molecules to dissociate before the surface has become saturated with other molecules can lead to higher incorporation levels than seen in the anneal case, in which molecules often have to desorb or migrate before BH$_2$ molecules have sufficient room to further dissociate. 
While temperatures during dosing should be limited to below at least \SI{450}{\celsius} to keep the hydrogen resist intact,\cite{vskerevn2020bipolar,Oura1990} our analysis indicates that dose temperatures closer toward this limit may be an effective tactic for increased incorporation levels. 
In contrast, the system has essentially no incorporation for anneal temperatures below \SI{200}{\celsius}. 
At around \SI{250}{\celsius}, we predict that incorporation rates will begin to increase as a function of temperature until hitting a saturation point around \SI{550}{\celsius}, again likely due to activating the B1 to C1 and B5 to C2 reactions. 
At higher anneal temperatures, the activated sheet density remains largely constant.

\v{S}kere\v{n} \textit{et al.}, in contrast, find a drastic increase in sheet density of boron after \SI{800}{\celsius}, reaching a maximum of \SI{4e14}{cm^{-2}}.\cite{vskerevn2020bipolar} They attribute this growth to the breaking apart of boron dimers from configurations such as B7, a process we notably do not model. At temperatures below \SI{800}{\celsius}, however, they measure an essentially constant sheet density of roughly \SI{1.0e14}{cm^{-2}}, comparing favorably with the \SI{0.84e14}{cm^{-2}} sheet density we predict at \SI{650}{\celsius}. This helps validate our assumption that bridging BH molecules are a reasonable proxy for final incorporation of a boron atom. 

This incorporated sheet density of activated boron at anneal temperatures below \SI{800}{\celsius} is roughly half of the density that has been seen from the analogous phosphine process ($\sim$\SI{1.6e14}{cm^{-2}}),\cite{simmons2005scanning} which can be achieved at considerably lower temperatures. 
This decrease in doping density can be directly attributed to the dimer nature of diborane. 
If we include boron atoms that end in a dimerized state (configuration B7 in Fig.~\ref{fig:rxn_pathway_schematic}) as activated boron atoms within our simulation, using just the \SI{120}{\celsius} dose and \SI{410}{\celsius} anneal, our predicted sheet density nearly triples to $\sim$\SI{2.4e14}{cm^{-2}}. 
The majority of boron deposited on the silicon surface by diborane, in fact, are therefore locked in electrically inactive configurations. 
The dimerization of boron atoms thus significantly decreases the maximal sheet density that can be achieved. 
While these dimerized boron atoms can be split apart by sufficient heating, it requires a significant amount of energy, occurring only at temperatures above \SI{800}{\celsius}. 
This thermal budget is undesirable and can lead to significant migration that degrades the precision nature of the doping process, particularly if the process is utilized in combination with a donor layer for complimentary logic.\cite{keizer2015suppressing} 
\begin{figure}
\includegraphics[width=0.6\columnwidth]{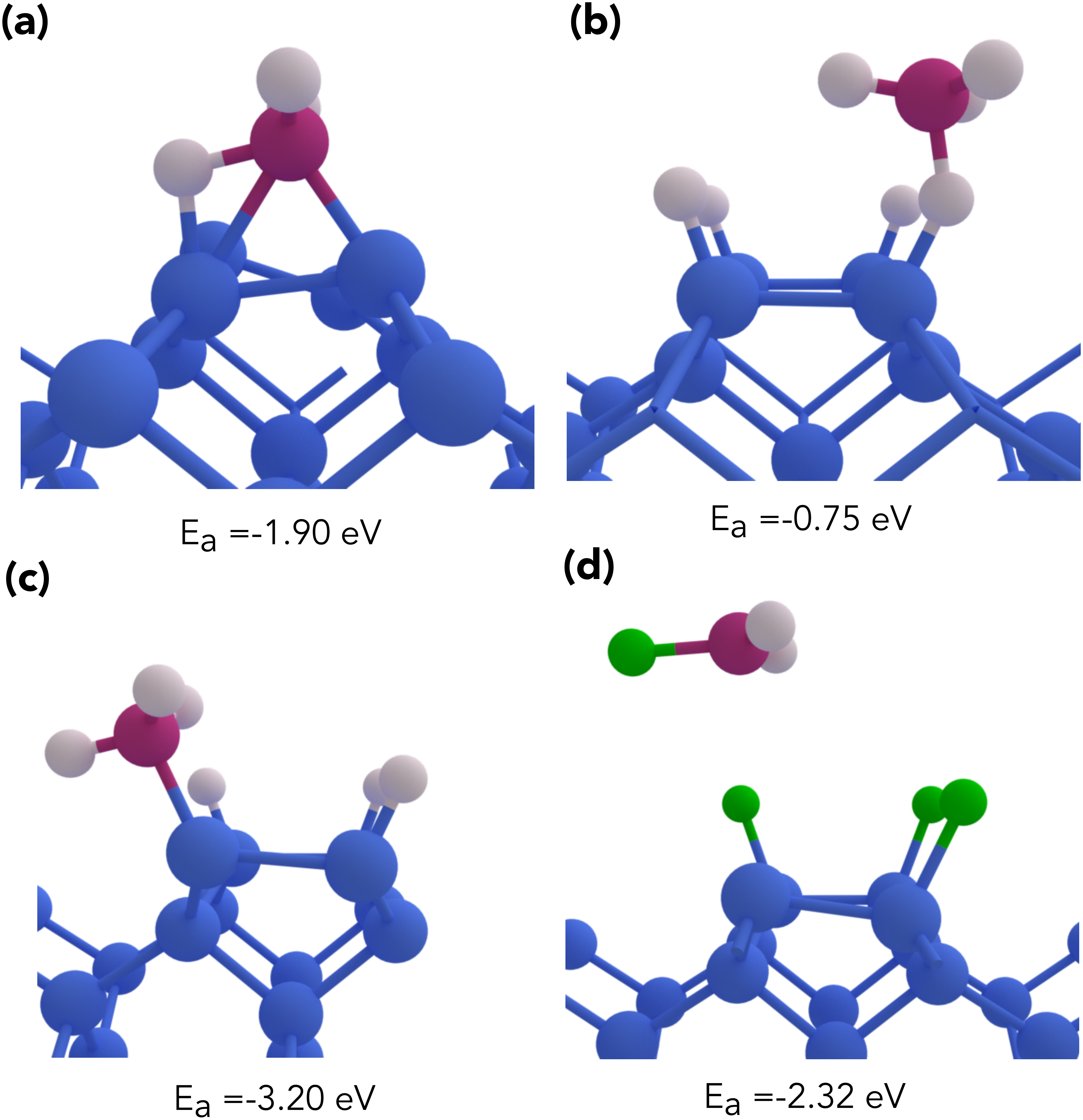}
\caption{ Side views of (a) BH$_3$ adsorbed onto a bare silicon surface, (b) BH$_3$ adsorbed onto a hydrogen terminated silicon surface, (c) BH$_2$ adsorbed onto a hydrogen terminated silicon surface, displacing the hydrogen and forming BH$_3$, and (d) BH$_2$ attempting to adsorb onto a chlorine terminated silicon surface, stripping the chlorine atom from the silicon and forming BH$_2$Cl. The significant adsorption energies for all these structures on both hydrogen and chlorine resists indicate that splitting diborane into BH$_x$ fragments to induce higher doping densities results in poor selectivity.
\label{fig:bhx_selectivity}}
\end{figure}

A conceivable alternative approach to increasing the final density of incorporated boron could be splitting diborane via heating into BH$_x$ fragments (here $1\leq x\leq5$) before exposing the resulting gas to silicon. While this would avoid the issue of seeding the surface with dimerized boron atoms that have a significant chance of remaining electrically inactive, we predict that the BH$_x$ fragments would have poor selectivity on bare silicon versus an atomic resist, ruining the atomic precision of dopant placement. In comparison to diborane, we predict BH$_3$ has a much stronger adsorption energy on bare silicon of --1.9 eV, shown in Fig.~\ref{fig:bhx_selectivity}a, matching the --1.86 eV reported by Konecny and Doren well.\cite{konecny1997adsorption} But BH$_3$ also has a relatively high adsorption energy on a hydrogen resist of --0.75 eV, as shown in Fig.~\ref{fig:bhx_selectivity}b. This indicates that while BH$_3$ will preferentially adsorb onto bare silicon, if it lands on hydrogen it will fully chemisorb onto the surface. The selectivity picture is even worse for BH$_2$: BH$_2$ has an adsorption energy of --3.2 eV on a hydrogen resist, stripping the hydrogen from the surface to form a BH$_3$ molecule that then directly attaches to the surface, as displayed in Fig.~\ref{fig:bhx_selectivity}c. Furthermore, the selectivity woes of BH$_2$ can not be mitigated by changing the atomic resist. For a chlorine resist, BH$_2$ has an adsorption energy of --2.32 eV, stripping a chlorine atom from the resist to form BH$_2$Cl, with a structure shown in Fig.~\ref{fig:bhx_selectivity}d. This leaves a bare silicon site exposed for the next BH$_x$ fragment to directly attach to. The selectivity of BH$_2$ on both hydrogen and chlorine is poor enough to make the splitting of diborane before exposure to silicon a likely non-viable option.

Overall, our results highlight how the dimer nature of diborane limits its capabilities as an APAM precursor. The final rate of dopant incorporation in small dimer windows is notably less than in phosphine and can be highly controlled by artificially increasing or decreasing the likelihood of diborane splitting into two BH$_3$ molecules on separate dimers, revealing the importance of this split to final incorporation. The high reaction barriers of the diborane dissociation pathway force relatively high operating temperatures, and even at an anneal temperature of \SI{650}{\celsius}, diborane produces an activated sheet density that is half that seen in phosphine at lower temperatures. The incorporated sheet density triples when electrically inactive, dimerized boron atoms are counted, indicating that the majority of boron deposited on the surface is locked in an electrically inactive state. Furthermore, this problem cannot be easily overcome by heating to split the boron dimers, either before or after exposure to silicon. Our DFT calculations predict that splitting diborane into BH$_x$ fragments before exposure to silicon leads to a lack of selectivity with both hydrogen and chlorine resists, and \v{S}kere\v{n} \textit{et al.} show that temperatures in excess of \SI{800}{\celsius} are required to split boron dimers once they have been adsorbed onto the silicon surface, a temperature that produces undesirable migration effects in the larger silicon device. This work therefore suggests that while diborane can work as an atomic precision acceptor precursor, it has inherent limitations. Alternative precursor molecules that do not naturally dimerize, such as BCl$_3$, will likely yield improved APAM performance. 

\section{Conclusion}
\label{sec:conclusion}
We have performed a detailed analysis of the dissociation pathway of diborane on the Si(100)-2x1 surface, calculating both adsorption energies and reaction barriers.
We find that diborane must overcome a barrier to adsorb onto the bare silicon surface, explaining the experimentally observed need to heat the sample to achieve significant levels of diborane adsorption. 
We then used our dissociation pathway in a KMC model to determine incorporation statistics for both small and large windows in the resist.
We demonstrate stochastic incorporation of an acceptor into silicon in resist windows between three and five dimers wide. 
This incorporation rate can be heavily influenced, however, by small errors in the calculation of the reaction barrier for the initial splitting of diborane onto two separate dimers or by the initial loss of hydrogen from one of the BH$_3$ fragments. 
We then predict the temperature dependence of incorporation in larger depassivated windows. 
Due to the relatively high reaction barrier required for the development of a bridging BH molecule, elevated temperatures are favored in both the dose and anneal step of dissociation. 
We demonstrate that the dimer nature of diborane inherently limits its doping density in comparison to phosphine. 
Furthermore, splitting the diborane into BH$_x$ fragments before exposure to silicon is not a viable path as BH$_3$ and BH$_2$ exhibit poor selectivity for both hydrogen and chlorine resists. 
Our results highlight that the dimer nature of diborane inherently limits its potential as an acceptor precursor. 


\begin{acknowledgement}

We gratefully acknowledge Rick Muller and Peter Schultz for useful discussions.
This work was supported by the Laboratory Directed Research and Development program at Sandia National Laboratories under project 213017 (FAIR DEAL) and performed, in part, at the Center for Integrated Nanotechnologies, an Office of Science User Facility operated for the U.S. Deppartment of Energy (DOE) Office of Science. 
Sandia National Laboratories is a multi-mission laboratory managed and operated by National Technology and Engineering Solutions of Sandia, LLC, a wholly owned subsidiary of Honeywell International, Inc., for DOE’s National Nuclear Security Administration under contract DE-NA0003525.
This paper describes objective technical results and analysis.
Any subjective views or opinions that might be expressed in the paper do not necessarily represent the views of the U.S. Department of Energy or the United States Government.

\begin{table}[h]
\begin{center}
\begin{tabular}[b]{l c}
\hline
\textbf{Numerical simulations} & \\
\hline
Total Kernel Hours [$\mathrm{h}$]& 615100\\
Thermal Design Power Per Kernel [$\mathrm{W}$]& 5.75\\
Total Energy Consumption Simulations [$\mathrm{kWh}$] & 3536\\
Average Emission Of CO$_2$ &  \\
In New Mexico, USA [$\mathrm{kg/kWh}$]& 0.5644\\
Total CO$_2$-Emission For Numerical Simulations [$\mathrm{kg}$] & 1996\\
Were The Emissions Offset? & \textbf{No}\\
\hline
\textbf{Transport} & \\
\hline
Total CO$_2$-Emission For Transport [$\mathrm{kg}$] & 0\\
Were The Emissions Offset? & n/a\\
\hline
Total CO$_2$-Emission [$\mathrm{kg}$] & 1996\\
\hline
\hline

\end{tabular}

\end{center}
\caption{Carbon emissions involved in the DFT calculations to make this paper. Here we count the time of all DFT calculations for this project, even if the final results were not included in this work. Estimations have been calculated using the examples of Scientific CO$_2$nduct\cite{conduct} and are correct to the best of our knowledge.}
\end{table}

\end{acknowledgement}

\bibliography{library,extrabib}

\end{document}